\documentclass[prl,
                twocolumn,
                preprintnumbers,
                amsmath,
                amssymb,
                showpacs,
                superscriptaddress]{revtex4-1}
\usepackage{graphicx} 
\usepackage{dcolumn}  
\usepackage{bm}       
\usepackage{amsfonts}

\usepackage{amssymb}
\usepackage{xcolor}
\usepackage{soul}
\usepackage{braket}
\usepackage{amsmath}

\newcommand{\eq}[1]{Eq.~(\ref{#1})}
\newcommand{\ful}{\mbox{C$_{60}$}}

\begin{document}

\title{Ultrafast transfer and transient entrapment of photoexcited Mg electron in Mg@$\ful$}

\author{Mohamed El-Amine Madjet}
\email{m.madjet@gmail.com}
\affiliation{Department of Natural Sciences, D.L. Hubbard Center for Innovation, Northwest Missouri State University, Maryville, Missouri
64468, USA}

\author{Esam Ali}
\affiliation{Department of Natural Sciences, D.L. Hubbard Center for Innovation, Northwest Missouri State University, Maryville, Missouri
64468, USA}

\author{Marcelo Carignano}
\affiliation{Department of Biomedical Engineering, Northwestern University, Evanston, IL 60208, USA}

\author{Oriol Vendrell}
\affiliation{Theoretical Chemistry, Institute of Physical Chemistry \& Centre for Advanced Materials,
Heidelberg University, Im Neuenheimer Feld 229 \& 225, 69120 Heidelberg,Germany}

\author{Himadri S. Chakraborty}
\email{himadri@nwmissouri.edu}
\affiliation{Department of Natural Sciences, D.L. Hubbard Center for Innovation, Northwest Missouri State University, Maryville, Missouri
64468, USA}

\begin{abstract}
{Electron relaxation is studied in endofullerene Mg@$\ful$, after an initial localized photoexcitation in Mg, by nonadiabtic molecular dynamics simulations. To ensure reliability, two methods are used: i) an independent particle approach with a DFT description of the ground state and ii) HF ground state with many-body effects for the excited state dynamics. Both methods exhibit similar relaxation times leading to an ultrafast decay and charge transfer from Mg to $\ful$ within tens of femtoseconds. Method (i) further elicits a robust transient-trap of the transferred electron that can delay the electron-hole recombination. Results shall motivate experiments to probe these ultrafast processes by two-photon transient absorption spectroscopy in gas phase, in solution, or as thin films.}

\end{abstract}

\maketitle
Synthesis, extraction and isolation methods of endofullerenes with encapsulated atoms/molecules are fast developing~\cite{popov17,liu14}. Time domain spectroscopy of these stable, symmetric systems can test advances in laboratory techniques and access real time novel processes of fundamental and applied interest. Due to their exceptional properties, progression of technology piggybacks fullerene and endofullerene materials through applications in molecular devices~\cite{chandler19,chai20}, energy storage~\cite{friedl18} and conversion\cite{jeon19,collavini18}. 

Photoinduced charge transfer (CT) is a key process in organic photovoltaics whose donor-acceptor complexes are predominantly based on fullerene materials. This is because a fullerene molecule can be chemically tuned by choosing its endohedral core~\cite{ross09} or exohedral ligands including polymers~\cite{he11,li12} to control light absorption efficiency and carrier transport. Upon absorbing a photon, the energy converts to an exciton that either dissociates into free carriers or recombines depending on the electron-hole separation and excitonic binding energy. Of course, the dissociation is preferred for photovoltaics~\cite{emmerich20}. Thus, the decay and transfer of a ``hot" electron from one location of the molecular material to another is a fundamental sub-process of this mechanism~\cite{sato18,ortiz17,boschetto18,juvenal19,cheng19}. Therefore, gaining insights into the CT dynamics by addressing a simpler prototype system is very important.

These ultrafast processes occur on the femtoseconds (fs) to picoseconds (ps) time-scale and are driven by the strong coupling between ionic and electronic degrees of freedom. Frameworks based on nonadiabatic molecular dynamics (NAMD), therefore, are appropriate for providing accurate, comprehensive descriptions of the processes~\cite{sato18,nelson20}. A powerful experimental technique to probe such dynamics is the ultrafast transient absorption spectroscopy (UTAS)~\cite{berera09} using fs pulses or, more recently, attosecond pulses for greater resolution~\cite{driver20,geneaux19}. Indeed, photoinduced charge migration has been measured in the time domain for fullerene-based polymerized films~\cite{juvenal19} and heterojunctions~\cite{cheng19}, and also for bulks~\cite{he18} and nanorods~\cite{kedawat19}. However, these systems are large and, consequently, the relaxation pathways may intermix with concurrent processes that can wash out, mask or camouflage spectral information on fundamental CT mechanisms, including access to prominent transient events.
\begin{figure}[h!]
\includegraphics[width=8.5 cm]{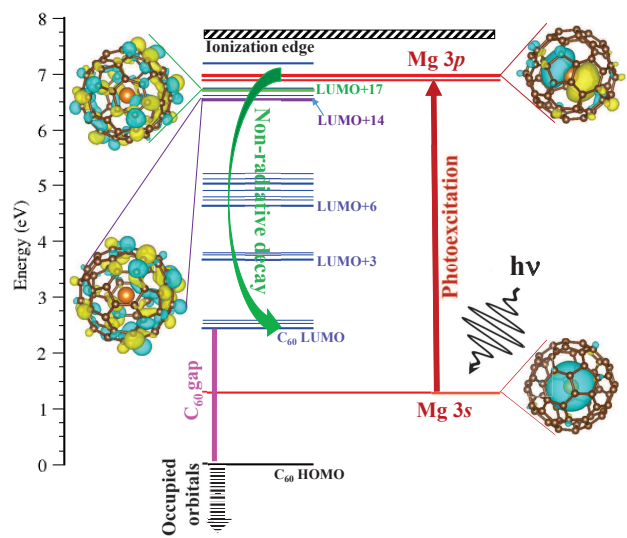}
\caption{(Color online) Mg@$\ful$ molecular orbital energies (relative to $\ful$ HOMO) at the DFT/B3LYP level of theory (see text). The Mg 3$s$ $\rightarrow$ 3$p$ photoexcitation and subsequent CT decay are illustrated. Isosurface plots of Mg 3$s$, LUMO+20 (Mg $3p$-type), LUMO+17 and LUMO+14 are shown.}
\label{fig1}
\end{figure}

\begin{figure}[h!]
\includegraphics[width=8.5 cm]{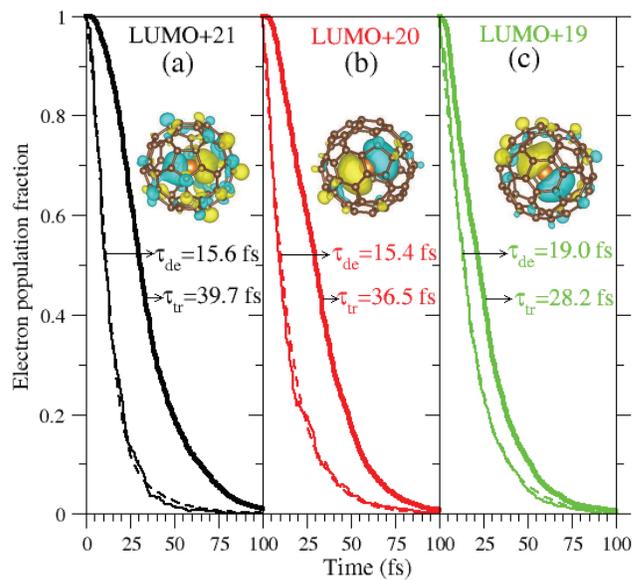}
\caption{Time evolutions of the decay and transfer population fractions after three initial excitations to LUMO+21 (a), LUMO+20 (b) and LUMO+19 (c), corresponding to Mg 3$p$ 3-fold degenerate orbitals. The decay ($\tau_{de}$) and transfer ($\tau_{tr}$) times shown are extracted by curve fittings (see text) and the fit curves for the decay are shown by the dashed lines.}
\label{fig2}
\end{figure}

From both theoretical and experimental standpoints, Mg@$\ful$ serves as an excellent benchmark system to study the ultrafast relaxation and charge separation of an exciton in contact with an organic matrix. (i) It features a ``surgical" photoexcitation at a local site.
(ii) It showcases a pristine relaxation dynamics, consisting of the photoelectron's transfer to an entirely different site. (iii) Its dynamics upon photoexcitation proceeds through potentially long-lived intermediate states, similarly as in transient charge-trappings~\cite{carey09}. Fig.\,\ref{fig1} delineates these points and displays the molecular orbital energies of Mg@$\ful$ as obtained in the present DFT description. Notice how the valence Mg 3$s$ level occurs isolated within the $\ful$ band gap and thus can be conveniently excited by a UV pump pulse to Mg 3$p$ - a level which splits into LUMO+19 to LUMO+21, each retaining predominant Mg character (Fig.\,\ref{fig2}). This happens owing to the lifting of the three-fold degeneracy of $p$ states due to the symmetry-breaking interaction with $\ful$. These excited states, localized on Mg, are the initial states of our simulation. Nonradiative decay, driven by electron-phonon couplings, then becomes the dominant decay process; no intercoulombic decay (ICD) channel~\cite{obaid20,javani14} exists, since the Mg excitation energy is lower than the $\ful$ ionization energy (IE).  This ultrafast relaxation is the subject of our simulations and can be followed by a time-delayed probe pulse in UTAS. As the photoelectron decays to LUMO+17, the first pure $\ful$ state (Fig.\,\ref{fig1}), an atom-to-$\ful$ CT occurs. This CT is complete and irreversible, and thus offers a clean and well defined event for experimental measurements. Decaying further, the electron lands on LUMO+14 and experiences a transient hold-up due to a wide energy gap right below this state which hinders the subsequent decay.  Consequently, the lifetime of the LUMO+14 population is predicted to be longer than for nearby states. Even though there are gaps below LUMO+6 and LUMO+3 in Fig.\,\ref{fig1}, their peak populations never grow enough due to significant slowdown.

We describe first the simulations performed with an independent particle (IP), molecular orbital description of the electrons. The ground-state geometry optimization of Mg@C$_{60}$ is conducted at the B3LYP/6-311+G$^{**}$ level of theory using Gamess~\cite{gamess1,gamess2}. For the empty $\ful$ optimized structure, this produces an accurate description of the band gap, 2.72 eV, close to reference values for molecular $\ful$~\cite{vinit17,zhang14}. Moreover, the calculated difference of 5.15 eV between the $\ful$ IE and electron affinity closely agrees with the difference of these quantities measured, respectively, by electron impact mass spectrometry~\cite{muigg96} and high-resolution photoelectron imaging~\cite{huang14}. Accounting for the known up-shift of DFT energies~\cite{schmidt15}, our computed free Mg IE matches the NIST value of 7.65 eV. This energy being close to the $\ful$ IE ensures that Mg 3$s$ in Mg@$\ful$ moves up into the $\ful$ band gap (Fig.\,\ref{fig1}) from higher screening by $\ful$ electrons.

\begin{figure}[h!]
\includegraphics*[width=9.0 cm] {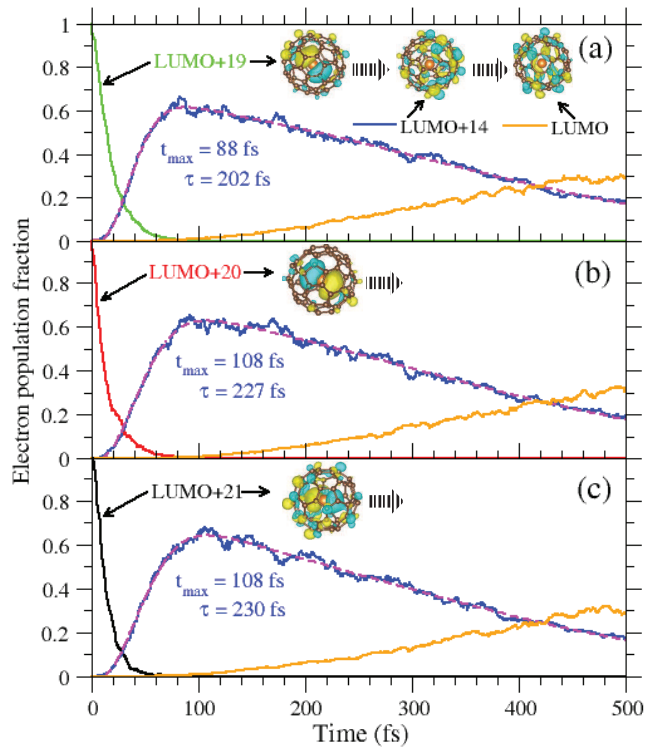}
\caption{(Color online) Time evolutions of the population fractions of the trapper state LUMO+14 and the band edge LUMO after initial excitations to LUMO+19 (a), LUMO+20 (b) and LUMO+21 (c). The excited state decays are also shown. The lifetime ($\tau$) and the time ($t_{max}$) of maximum population of LUMO+14 are extracted by curve fittings; the fit curves are shown as the dashed lines.} 
\label{fig3}
\end{figure}

From the optimized structure of Mg@$\ful$, we conduct MD simulations in the NVT ensemble at 300 K with a velocity-rescaling thermostat. Isosurface plots in Figs.\,\ref{fig1}-\ref{fig3} are obtained from the final structure of the NVT equilibration run. A production run starts in the NVE ensemble and extends to 3 ps. All MD simulations are conducted using the CP2K~\cite{cp2k} and B3LYP hybrid functional. The time-dependent population is obtained by averaging over 20 initial configurations and 1000 surface hopping trajectories for each configuration. The DFT-D3 dispersion correction of Grimme is employed to account for the dispersion interactions~\cite{Grimme1,Grimme2}. The QMflows-namd~\cite{qmflows} module interfaced with CP2K is employed to compute electronic structure properties ($\phi_j, \epsilon_j$) and to obtain electron-phonon nonadiabatic couplings (NACs) $d_{jk}$~\cite{guo18},
\begin{equation}\label{nacs}
d_{jk} = \frac{\left<\phi_j|\vec{\nabla}_{\!\!\scriptsize R} H|\phi_k\right>}{\epsilon_k-\epsilon_j}\frac{\partial\vec{R}}{\partial t},
\end{equation}
where $H$ and $\vec{R}$ are the electronic Hamiltonian and nuclear coordinate. Evidently, NACs can enhance by (i) larger orbital overlaps, (ii) narrower energy separations, and (iii) faster nuclear velocities. The energies and NACs are then used to perform the NAMD simulations using the PYXAID package; for details, see Ref.\,\cite{akimov1,akimov2,madjetccp,madjetjpcl}.

The simulations of the dynamics of Mg@$\ful$ start from a localized excitation of Mg 3$s$ to each of the 3$p$ states, LUMO+21, LUMO+20, and LUMP+19, by photon energies of 5.69 eV, 5.67 eV, and 5.59 eV respectively. Fig.\,\ref{fig2} presents the time evolution of the relaxations. From the initial excited state, the hot electron quickly spreads to the other two states of the 3$p$ degeneracy owing to their strong NACs from large orbital overlaps [\eq{nacs}] before the electron transfers to lower states. The time evolution of the population fractions of the initial states are plotted in the panels of Fig.\,\ref{fig2}. Their decay times ($\tau_{de} $) are evaluated by fitting to the sum of an exponential and a Gaussian decay function, as 15.6 fs, 15.4 fs and 19.0 fs respectively. $\tau_{de}$ for LUMO+19 is a bit longer because in this case the initial sputter of population to LUMO+20 and LUMO+21 turns around to feed LUMO+19 back. 

While LUMO+17 is found to be the first dominant $\ful$ state on the decay path, LUMO+18 is an atom-$\ful$ hybrid state. Thus, in order to estimate the atom to $\ful$ electron transfer time ($\tau_{tr}$) from a given initial excited state, we add up the population of the three 3$p$ states and half of that of hybrid LUMO+18. The resultant cumulative curves, representing the transfer dynamics, are also included in Fig.\,\ref{fig2}. Fittings yield the values of $\tau_{tr}$ to be 39.7 fs, 36.5 fs, and 28.2 fs respectively. The slower transfer trend going from the higher to lower initial state points to the fact that the higher the excitation the longer the electron takes to evacuate the Mg region. Further, one may visualize the original electron-hole pair in Mg as a {\em local} exciton, while the exciton after the electron transfers to the cage with a hole at Mg 3$s$ is a {\em nonlocal} exciton. Therefore, $\tau_{tr}$ also corresponds to the ultrafast conversion time from a local to a nonlocal exciton, leading to the generation of carriers. Vibronic coupling suppresses the electron-hole recombination owing to energy dissipation to the vibrational modes of {\ful}. In solid $\ful$, the exciton biding energy has a large value $\sim$ 0.5 eV \cite{golden95}, which disfavors their separation to free charges.

The state LUMO+14 acts as a transient trapper for the electron. Indeed, based on \eq{nacs}, the population of LUMO+14 grows at a higher rate fed by energetically close states above it (Fig.\,\ref{fig1}), while the energy gap below LUMO+14 leads to a slower depopulation. Panels of Fig.\,\ref{fig3} show the net electron population dynamics of LUMO+14 for each initial Mg excitation. For all cases, the peak of LUMO+14 rises to about 65\% of the population. Hence, this transient population can likely be experimentally accessed. A sum of exponential and Gaussian growth {\em plus} decay functions are used to fit the curves to extract the trapping times ($\tau$) of 202 fs, 227 fs and 230 fs respectively. Again, the electron residing somewhat longer in the atomic zone when excited to a higher state feeds LUMO+14 for a longer time. This fact is reflected from an earlier time ($t_{max}$) of 88 fs at the maximum population of LUMO+14 when the electron was initially excited to LUMO+19. The similar $t_{max}$ values of about 108 fs for the other two cases occur likely because of the complexity of the dynamics. Only after 100 fs the LUMO starts to gain population and reaches about 30\% at 500 fs (Fig.\,\ref{fig3}). This significant slowdown at the band edge is due to additional slowing effects induced by the gaps below LUMO+6 and LUMO+3. However, like the population peak of LUMO, that of LUMO+6 and LUMO+3 are so low that it will likely be difficult to measure them. We note that such intermittent gaps in $\ful$ unoccupied levels were found in other calculations~\cite{schmidt15}. Thus, with the reliability of the B3LYP functional, the prediction of a strong population trap atop the first gap on the decay path must be realistic.

The computation scheme used above in the IP framework neglected the electron-electron and electron-hole interactions. To account for the many-electron effects, we now apply a method where the dynamics of excited (many-electron) states are computed using on-the-fly NAMD simulations as implemented in our CDKT~\cite{madjet13} tool interfaced with Gamess-US~\cite{gamess1,gamess2}. Excited state energies and gradients are calculated at the CIS/SBKJC level analytically as implemented in Gamess-US. The non-adiabatic couplings are described by a Landau-Zener (LZ) approximation~\cite{madjet13}. The computational costs are further reduced by using an
effective core potential (ECP) basis set, namely SBKJC~\cite{sbkj}, instead of an all-electron basis. The accuracy of the SBKJC basis set is ascertained by comparing the excited state energies and oscillator strengths with those calculated using the all-electron basis set 6-31G for the first 200 singlet excited states and also using hybrid functional PBE0 in TDDFT.  

\begin{figure}
\includegraphics[width=8.5 cm, keepaspectratio]{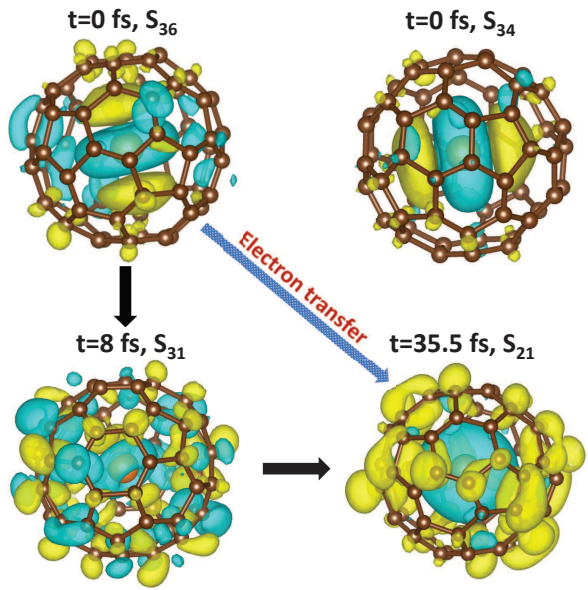}
\caption{(Color online) Electron density difference (excited {\em minus} ground) of Mg@$\ful$ with iso-density value of $\pm$ 0.0005 a.u.\ with positive (golden) and negative (green) values. Results at $t=0$ for S$_{36}$ and S$_{34}$, and the electron transfer for S$_{36}$ in a molecular dynamics trajectory are shown.}
\label{fig4}
\end{figure}

The insertion of Mg inside $\ful$ induces a blue shift of the bright states of empty $\ful$. For the optimized structure, the excitation of Mg corresponds to an energy of 4.95 eV and has an oscillator strength of $\sim$ 0.15 (allowed transition). The corresponding excited state is three fold degenerate, as expected, with energies close to the 3$s$ $\rightarrow$ LUMO+21, LUMO+20, LUMO+19 excitations in the IP treatment, and corresponds to the CIS (many electron) excited states: S$_{34}$, S$_{35}$ and S$_{36}$. Fig.\,\ref{fig4} shows the electron density difference, excited {\em minus} ground, for S$_{36}$ and S$_{34}$ at $t=0$, where their localized Mg nature can be seen; the golden color encodes positive values while the green is negative. The central negative lobes are due to the subtraction of the ground Mg 3$s$ spherical density.   

Thirty nine singlet states (S$_{1}$-S$_{39}$) are included in the simulation and the classical equations of motion are integrated with the velocity Verlet algorithm. The trajectory is propagated for 150 fs with
time-steps of 0.5~fs for the vibrational dynamics. In our implementation of surface-hopping based on the LZ model, the momentum of the nuclei is re-scaled along the direction of the gradient difference vector of the potential energy surfaces involved in the transition to ensure the conservation of total energy. After the initial ($t=0$) excitation of Mg, as illustrated in Fig.\,\ref{fig4} for S$_{36}$ in a NAMD trajectory, the electron population follows structural rearrangements driven non-adiabatically by the electron-phonon couplings. After 35.5 fs, S$_{36}$ decays to S$_{21}$ to complete the electron transfer to $\ful$. This is evident in the iso-density plot of S$_{21}$ where the negative values are localized on Mg and the positive values are on the cage. Note also the plot at an intermediate time of 8 fs which denotes a hybrid state. The 35.5-fs transfer time is very close to the average of three values of $\tau_{ tr}$ in Fig.\,\ref{fig2}. This suggests that the many-body dynamics, which dominates the plasmon-driven ionization spectra~\cite{chakraborty08} at higher energies (XUV), is not important in the middle UV region of current interest. Furthermore, most of the excited states below S$_{34}$ are found to be dark -- a fact that favors a localized photoexcitation of Mg -- and are only populated during the relaxation of the hot electron.

Similar to our IP model, there are energy gaps between CIS excited states below S$_{21}$. These gaps ($\sim$ 0.3 eV) are not large enough to result in a transient capture of the excited population, which now continues to relax towards lower-energy states. A reason for such denser CIS spectrum is the presence of satellite states originating from linear combinations of coupled particle-hole configurations. However, whether this higher density of states necessarily quenches the trapping mechanism observed in the IP model remains an open question. The current study combines CIS trajectories with a LZ scheme to calculate the hopping probability. But it is conceivable that some of the NACs computed via Eq.~(\ref{nacs}) (using electronic states instead of orbitals) would be smaller than the LZ prediction (two states may feature a small energy gap, yet the numerator in Eq.~(\ref{nacs}) may be small or even vanish). However, the evaluation of accurate NACs for correlated states in this large system is currently out of reach. In order to draw a fairer comparison, one also needs to perform B3LYP/TDDFT calculations, besides HF/CIS, to bring the correlated calculations to same footing as the IP method. On the other hand, the excellent match between the CT times by both methods suggests that many-electron effects may be weak at the onset of relaxation. The differences found in the subsequent dynamics, as to whether a transient trapping of the electronic population might be present, offers a unique motivation to conduct experiments and extend calculations in endofullerenes.  

One way to synthesize Mg@$\ful$ for the experiment is by the ion implantation technique. $\ful$ films can be irradiated by Mg ions~\cite{lee20} in a similar manner employed for Li@$\ful$, which showed stability in the air after sublimation~\cite{campbell97}. The ion energy can be optimized to allow the encapsulation and yet to minimize the destruction of fullerenes, so Mg@$\ful$ can be isolated from the collision debris. The air stability of the molecule will be a challenge owing to the oxidation state Mg$^{2+}$@C$_{60}^{2-}$. This can be mitigated by converting to a stable salt-form with Mg$^{2+}$@$\ful$ cation and some stabilizing anion, as was accomplished for Li@$\ful$~\cite{aoyagi10}. For example, Ca@$\ful$ was produced with a laser vaporization source and its photoelectron spectroscopy in gas-phase was performed~\cite{wang1993}. Due to the non-covalent interactions of Mg@$\ful$ with its environment, we believe that the essence of our results will remain valid also in solution or in thin films.

To conclude, we simulated and analyzed the ultrafast nonradiative relaxation process, driven by electron-phonon coupling (lattice thermalization), of a photoexcited hot electron in an atom confined in $\ful$. Mg@$\ful$ features a clean and uncluttered electron relaxation to the outer {\ful} shell, possibly featuring a transient slow-down of the electron relaxation process in real time due to the presence of large gaps in the spectrum of excited electronic states. The possibility of inducing the initial excitation accurately within Mg makes this molecule an ideal example for ultrafast transient absorption spectroscopic studies. Good agreement at early times between the two employed methods, with and without the many-body interactions, indicates that the ultrafast charge separation of the initial exciton with tens of femtoseconds driven by vibronic effects is a robust result. The study provides a reference to understand both experimental and theoretical investigations on endofullerene derivatives with increasing structural complications {\em via} functionalization and we hope that the current research will motivate experimental activities in the domain of ultrafast science.

\begin{acknowledgments} 
Computing time at Bartik High-Performance Computing Cluster in Northwest Missouri State University is acknowledged. Dr.\ Felipe Zapata is acknowledged for help and assistance with Qmflows code. We thank Dr. Alexey Popov and Dr. Eleanor Campbell for encouraging discussions on synthesis possibilities of Mg@$\ful$ for future experiments. The research is supported by the National Science Foundation Grant No.\ PHY-1806206, USA. 
\end{acknowledgments}


\end{document}